\documentstyle[12pt]{article}
\begin{document}

\title{Supersymmetric Method for Constructing
Quasi-Exactly and Conditionally-Exactly Solvable Potentials}

  \author{V. M. Tkachuk\\ 
  {\small Ivan Franko Lviv State University, Chair of Theoretical Physics }\\ 
	{\small 12 Drahomanov Str., Lviv UA--290005, Ukraine}\\
	{\small E-mail: tkachuk@ktf.franko.lviv.ua}}

\maketitle

\begin{abstract}
Using supersymmetric quantum mechanics we develop a new me\-thod
for constructing quasi-exactly solvable (QES) potentials 
with two known eigenstates.
This method is extended for constructing conditionally-exactly
solvable potentials (CES).
The considered
QES potentials at certain values of parameters become 
exactly solvable and can be treated as CES ones.

Keywords: supersymmetry, quantum mechanics,
quasi-exactly solvable potentials, 
conditionally-exactly solvable potentials.

PACS numbers: 03.65.-w; 11.30.Pb.
\end{abstract}
\pagebreak

\section{Introduction}
Since the appearance of quantum mechanics there has been
continual interest in models for which the corresponding Schr\"{o}dinger
equation is exactly solvable. With regards to
solvability of the Schr\"{o}dinger equation 
there are three interesting classes
of the potentials.

The first class is the exactly solvable potentials
allowing to obtain in explicit form all energy levels
and corresponding wave functions.
The hydrogen atom and harmonic oscillator are the best-known
examples of this type.

The second class is the so-called quasi-exactly solvable (QES)
potentials for which a finite number of eigenstates of corresponding 
Hamiltonian can be found exactly in explicit form.
The first examples of QES potentials were given in [1--4].
Subsequently several methods for generating QES potentials were worked out
and as a result many QES potentials were found [5--13] 
(see also review book \cite{14}). 
Three different methods that are based respectively
on the polynomial ansatz for wave functions, the point canonical
transformation, and the supersymmetric (SUSY) quantum mechanics are 
described in \cite{12}.
Recently, an anti-isospectral transformation
called also as duality transformation was introduced in \cite{15}. 
This transformation
relates the energy levels and wave functions of two QES potentials.
In \cite{16} a new QES potential was discovered using this 
anti-isospectral transformation.

The third class is the conditionally-exactly solvable (CES)
potentials for which the eigenvalues problem
for the corresponding Hamiltonian is exactly solvable only when 
the parameters of the potential obey certain conditions.
Such a class of potentials was first considered in \cite{21}.
It is interesting to note that
in \cite{Nag} it was demonstrated the equivalence of the condition
required for potential obtained in \cite{21} to be a CES
potential with the condition that this potential can be put in
an explicitly supersymmetric form.
Recently, new examples of CES potentials have been discovered 
\cite{Dutt,19,20}.

Very useful algebraic tool for investigation of the problem
of exact solvability of the Schr\"{o}dinger equation is the SUSY
quantum mechanics (for review of SUSY quantum mechanics see \cite{17,18}).
For constructing QES potentials the SUSY method 
was used in [10--12]. The starting point of this
method is some initial QES potential with
$n+1$ known eigenstates. Then applying the technique of
SUSY quantum mechanics one can calculate the supersymmetric
partner of the QES potential which
is a new QES potential with $n$ known eigenstates.
In the papers \cite{19,20} the SUSY quantum mechanics was used to develop 
some generalized method
for the constructing the CES potentials.

In our previous paper \cite{25} we have proposed 
a new SUSY method for constructing
QES potentials with two known eigenstates which in contrast to
the papers [10--12] does not require knowledge of the initial QES
potentials for a generation of a new QES ones. 
In \cite{26} we extended this method for constructing QES potentials
with three known eigenstates.

The present paper is devoted to further development of the 
SUSY method proposed in \cite{25} and to extension of this method
for constructing CES potentials.
In result we obtain new QES and CES potentials. 
A new interesting point is that QES potentials
with two known eigenstates
become exactly solvable
at certain fixed values of parameter and therefore they can   
be treated as CES potentials.

\section{Supersymmetric quantum mechanics}

In the Witten's model of supersymmetric quantum mechanics
the SUSY partner Hamiltonians $H_\pm$ read
\begin{equation} \label{5}
H_\pm=B^\mp B^\pm=-{1\over2}{d^2\over dx^2}+ V_\pm(x),
\end{equation}
where
\begin{eqnarray} \label{2}
B^\pm={1\over\sqrt{2}}\left(\mp{d\over dx}+ W(x)\right), \\ \label{221}	
V_\pm (x)={1\over 2}\left(W^2(x) \pm W'(x)\right), \ \ W'(x)={dW(x)\over
dx},
\end{eqnarray}
$W(x)$ is referred to as a superpotential.
In this paper we shall consider the  systems 
on the full real line 
$-\infty<x<\infty$.

The eigenvalues $E^{\pm}_n$ and eigenfunctions $\psi^{\pm}_n(x)$
of the Hamiltonians $H_{\pm}$
are related by SUSY transformations
which in the case of unbroken SUSY read
\begin{eqnarray} 
&& E_{n+1}^-=E_n^+,\ \ E_0^-=0,  \label{11} \\
&& \psi_{n+1}^-(x)={1\over \sqrt{E_n^+}}B^+\psi_n^+(x), \label{12} \ \
\psi_n^+(x)={1\over \sqrt{E_{n+1}^-}}B^-\psi_{n+1}^-(x). 
\end{eqnarray}

As a consequence of SUSY 
the Hamiltonians $H_+$ and $H_-$ 
have the same energy spectrum
except for the zero energy ground state. 
The latter exists in the case of 
the unbroken SUSY. 
Only one of the Hamiltonians
$H_\pm$ has a square integrable eigenfunction corresponding to
the zero energy. Here we use the convention
that the zero energy eigenstate belongs to $H_-$.
Due to the
factorization of the Hamiltonians $H_\pm$ (see (\ref{5}))
the ground state for $H_-$ satisfies the equation
$
B^-\psi_0^-(x)=0
$
the solution of which is
\begin{equation} \label{9}
\psi_0^-(x)=C^-_0\ \exp\left(-\int W(x) dx\right),
\end{equation}  
$C^-_0$ is the normalization constant. Here and below $C$ denotes
the normalization constant of the corresponding wave function.
From the condition of square integrability of 
the wave function $\psi^-_0(x)$ 
it follows that the superpotential must satisfy the condition
\begin{equation} \label{10}
{\rm sign}(W(\pm\infty)) = \pm 1,
\end{equation}
which is the condition of the existence of unbroken SUSY.

For a detailed description of SUSY quantum mechanics and
its application for the exact calculation of eigenstates of Hamiltonians
see reviews \cite{17,18}.
Just the properties of the unbroken SUSY quantum mechanics 
reflected in SUSY transformation (\ref{11}), (\ref{12})
can be used for exact calculation of the energy 
spectrum and wave functions.
In the present paper we use these properties for the generation of the 
QES potentials with two known eigenstates and CES potentials.

\section{QES potentials with two known eigenstates} 

We shall solve the eigenvalue problem for the Hamiltonian $H_-$. 
The ground state of this Hamiltonian is known and is given by 
wave function (\ref{9}) with energy $E^-_0=0$.
In order to calculate the excited state of $H_-$ we use the following
well-known procedure used in SUSY quantum mechanics. 
Let us consider the SUSY partner of $H_-$, i.e. the Hamiltonian $H_+$. 
If we calculate the ground state of $H_+$ we immediately find the first 
excited state of $H_-$ using the SUSY 
transformations (\ref{11}), (\ref{12}). 
In order to calculate 
the ground state of $H_+$ let us rewrite it in the following form
\begin{equation} \label{14}
H_+=H_-^{(1)} + \epsilon = B_1^+B_1^- + \epsilon, \ \  \epsilon > 0,
\end {equation}
which leads to the following relation between the potential energies
\begin{equation}\label{171}
V_+(x)= V^{(1)}_-(x)+\epsilon,
\end{equation}
and superpotentials
\begin{equation} \label{19}
W^2(x)+W'(x)=W_1^2(x)-W'_1(x) +2 \epsilon,
\end{equation}
where 
$ \epsilon$ is the energy of the ground state of $H_+$ 
since we supposed that $H_-^{(1)}$ 
similarly to $H_-$
has zero energy ground state,
$B_1^{\pm}$ and $V^{(1)}_-(x)$ are given by (\ref{2}) and (\ref{221})
with the new superpotential $W_1(x)$.  

As we see from (\ref{14}) the ground state wave function
of $H_+$ is also the ground state wave function of $H_-^{(1)}$ 
and it satisfies the equation
$B^-_1\psi_0^+(x)=0$.
The solution of this equation is
\begin{equation} \label{17}
\psi_0^+(x)=C^+_0 \ \exp\left(-\int W_1(x) dx\right),
\end{equation}  
where for the square integrability of this function
the superpotential $W_1(x)$ must satisfy 
the same condition as $W(x)$ (\ref{10}). 
Using (\ref{11}) and (\ref{12}) we obtain the energy level
$E_1^-=\epsilon$
and the wave function of the first excited state 
$\psi_1^-(x)$  for $H_-$.

Repeating this procedure in the case of shape invariant 
potentials \cite{22} and self-similar potentials \cite{Shab,Spir} 
it is possible
to calculate all energy spectrum and the corresponding wave functions.
As a result for these cases many exactly solvable 
potentials were obtained \cite{SS}
(see also review \cite{17}).

We consider a more
general case and do not restrict ourselves to the 
shape invariant potentials or self-similar potentials.
In this case it is not possible
to obtain all energy spectrum.
In \cite{25} we obtained the general solution of equation
(\ref{19}) and thus derived a general expression for 
QES potential with two explicitly known eigenstates.
The basic idea consists in finding
such a pair of $W(x)$ and
$W_1(x)$ that satisfies equation (\ref{19}).
To do this we rewrite equation (\ref{19}) in the following form
\begin{equation} \label{20}
W'_+(x)=W_-(x)W_+(x) +2\epsilon,
\end{equation}
where
\begin{eqnarray} \label{21}
W_+(x)=W_1(x) + W(x),\\ \nonumber
W_-(x)=W_1(x) - W(x).
\end{eqnarray}
This new equation (\ref{20}) can be easily solved
with respect to
$W_-(x)$ for a given arbitrary function $W_+(x)$ 
or with respect to $W_+(x)$ for a given arbitrary function $W_-(x)$. 
Then from (\ref{21}) we obtain superpotentials $W(x)$ and $W_1(x)$ which
satisfy equation (\ref{19}).

In contrast to our paper \cite{25} where we use the solution
with respect to $W_-(x)$, in the present paper we use the solution
with respect to $W_+(x)$. 
This solution is more convenient
for construction CES potentials and can be written in 
the following form
\begin{equation}\label{29}
W_+(x)=\exp\left(
\int dx W_-(x)\right)\left[2\epsilon\int dx\exp\left(
-\int dx W_-(x)\right)+ \lambda \right],
\end{equation}
here $\lambda$ is the constant of integration.

In order to simplify solution (\ref{29}) let us choose
$W_-(x)$ to be of the form
\begin{equation}\label{30}
W_-(x)=-\phi''(x)/\phi'(x).
\end{equation}
To provide 
a nonsingularity of $W_-(x)$ and as a result nonsingularity of
$V_{\pm}(x)$
we shall consider a nonsingular
monotonic function $\phi(x)$ 
satisfying the condition $\phi'(x)>0$. 
Then
substituting (\ref{30}) into (\ref{29}) we obtain
\begin{equation}\label{31}
W_+(x)=(2\epsilon\phi(x)+\lambda)/\phi'(x).
\end{equation}\
Note that the constant $\lambda$ can be included into the function 
$\phi(x)$ and thus for $W_+(x)$ we obtain
\begin{equation} \label{32}
W_+(x)=2\epsilon\phi(x)/\phi'(x).
\end{equation}
Finally, for superpotentials $W(x)$ and $W_1(x)$ we have
\begin{eqnarray}\label{33}
W(x)=(\epsilon\phi(x)+{1\over2}\phi''(x))/\phi'(x), \\ 
W_1(x)=(\epsilon\phi(x)-{1\over2}\phi''(x))/\phi'(x). \label{34}
\end{eqnarray} 

Using this result for the wave functions of the ground state
with the energy $E^-_0=0$ and excited state with $E^-_1=\epsilon$
we obtain
\begin{eqnarray} \label{35}
\psi^-_0(x)=C^-_0(\phi'(x))^{-1/2}\exp\left(
-\epsilon\int dx \phi(x)/\phi'(x)\right), \ \ E^-_0=0,\\ \label{351}
\psi^-_1(x)=C^-_1\phi(x)(\phi'(x))^{-1/2}\exp\left(
-\epsilon\int dx \phi(x)/\phi'(x)\right), \ \ E^-_1=\epsilon.
\end{eqnarray}

Note that as we see from (\ref{21})
$W_+(x)$ must satisfy the same condition (\ref{10}) as 
$W(x)$ and $W_1(x)$ do. Then because $\phi(x)$ is monotonic
function and $\phi'(x)>0$ from (\ref{32}) it follows
that $\phi(x)$ has one node.
Therefore 
$\psi^-_1(x)$ given by (\ref{351})
also has one node and thus corresponds to the first excited state.
The functions $\phi(x)$ that satisfy described condition
provide also the square integrability of the wave 
functions (\ref{35}) and (\ref{351}).

It is worth to stress
that $\phi (x)=\psi^-_1(x)/\psi^-_0(x)$
from which follows an interesting fact.
Namely, the ratio $\phi(x)$ of the wave functions of first excited
state and ground state and the distance $\epsilon$ between
the corresponding energy levels entirely determine 
the potential energy.

QES potential $V_-(x)$ is given by (\ref{221}) 
with superpotential (\ref{33}).
Choosing different $\phi(x)$ and $\epsilon$ 
we obtain different QES potentials
with explicitly known two eigenstates. 

Now let us consider interesting examples of new QES potentials
which become exactly solvable at certain fixed 
values of parameter $\epsilon$
and thus can be treated as CES ones.

\subsection{Example 1}

Let us put
\begin{equation} \label{E1phi}
\phi (x)= \beta H_{2k+1}(ix),
\end{equation}
where $H_m(x)$ is Hermite polynomial, the final result does not depend 
on the constant $\beta$.
The superpotentials in this case read
\begin{eqnarray}
W(x)={\gamma}x+i2k\left({\gamma}+1 \right)
{H_{2k-1}(ix)\over H_{2k}(ix)}, \label{E1W} \\
W_1(x)={\gamma}x+i2k\left({\gamma}-1 \right)
{H_{2k-1}(ix)\over H_{2k}(ix)}, \label{E1W1}
\end{eqnarray}
where we have introduced the notation
\begin{equation}
\gamma ={\epsilon \over 2k+1}.
\end{equation}
Substituting the superpotential $W(x)$ (\ref{E1W})
into (\ref{221}) we obtain the following QES potential $V_-(x)$ 
\pagebreak
\begin{eqnarray} \label{E1V}
V_-(x)={1\over 2}{\gamma}^2 x^2+
2k(2k-1)\left({\gamma}+1\right)^2
{H_{2k-2}(ix)\over H_{2k}(ix)} \\ \nonumber
-2k^2\left({\gamma}+1\right)
\left({\gamma}+3\right)
\left({H_{2k-1}(ix)\over H_{2k}(ix)}\right)^2 \\ \nonumber
+k{\gamma}\left({\gamma}+1\right)-
{1\over 2}{\gamma}.
\end{eqnarray}
The wave functions of the ground and first excited
states read
\begin{eqnarray}
\psi_0^-(x)=C_0^- \left(H_{2k}(ix)\right)^{-(1+\gamma )/2}
\exp (-\gamma x^2 /2), \\
\psi_1^-(x)=C_1^-H_{2k+1}(ix) \left(H_{2k}(ix)\right)^{-(1+\gamma )/2}
\exp (-\gamma x^2 /2).
\end{eqnarray}

It is interesting to note that in the special case $\gamma =1$,
i.e. 
\begin{equation} \label{E1Eps}
\epsilon =2k+1,
\end{equation} 
the second term in $W_1(x)$ drops up
and $W_1(x)$ corresponds to the superpotential of a linear
harmonic oscillator. 
Then $V^{(1)}_-(x)$ and, as a result of (\ref{171}),
$V_+(x)$ are the potential energies of the linear harmonic oscillator.
Therefore in this case the SUSY partner
$H_+$ is the Hamiltonian of the linear harmonic oscillator and we know
all its eigenfunctions in explicit form.
Using SUSY transformations (\ref{11}), (\ref{12}) we can easily calculate
the energy levels and the wave functions of all the excited states of
$H_-$. 
The energy spectrum of $H_-$ in this special case is the following
\begin{equation}
E_0^-=0, \ \ E_n^-=n+2k, \  n=1, 2, ...
\end{equation}

Thus QES potential (\ref{E1V}) at fixed value of $\epsilon$ (\ref{E1Eps})
becomes exactly solvable 
and therefore can be treated as CES potential.
Note that $V_-(x)$ in this special case $\gamma=1$
corresponds to the potential 
obtained by Bagrov and Samsonov \cite{23,24} via Darboux method and 
latter by Junker and Roy \cite{19,20} within SUSY approach.

\subsection{Example 2}

Consider the function
\begin{equation}
\phi (x)= \beta {H_{2k+1}(ix)\over H_{2m}(ix)}, \ \ k\ge m,
\end{equation}
which generalizes the one given in the first example.
For superpotentials we obtain
\begin{eqnarray}
W(x)&=&- x-i4m{H_{2m-1}(ix)\over H_{2m}(ix)}   \nonumber\\
&&-i{\epsilon+2k-2m+1 \over H_{2m+1}(ix)/H_{2m}(ix)- 
H_{2k+2}(ix)/H_{2k+1}(ix)}, \label{E2W}  \\
W_1(x)&=&x+i4m{H_{2m-1}(ix)\over H_{2m}(ix)}  \nonumber \\
&&-i{\epsilon-(2k-2m+1)\over H_{2m+1}(ix)/H_{2m}(ix)-  
H_{2k+1}(ix)/H_{2k}(ix)}. \label{E2W1}
\end{eqnarray}
The QES potential $V_-(x)$ is given by (\ref{221}) with superpotential
(\ref{E2W}).
The expressions for the QES potential and the wave functions in
this case are somewhat complicated and we do not write them down
in explicit form.

We would like to stress the following very interesting point.
In the special case 
\begin{equation} \label{E2EPS}
\epsilon =2k-2m+1 
\end{equation}
the second term in
superpotential $W_1(x)$ (\ref{E2W1}) 
drops up and then this $W_1(x)$ coincides with
superpotential $W(x)$ (\ref{E1W}) from the first example 
for $\gamma =1$ which is exactly solvable.
Then using the same explanation as in the end of first example 
we may conclude that for the potential $V_-(x)$ calculated with
superpotential (\ref{E2W}) in the special case (\ref{E2EPS})
it is possible to get all energy levels and corresponding
wave functions and thus  $V_-(x)$ can be
treated as CES potential.
For the energy levels we obtain
\begin{eqnarray}
E_0^-=0,\ \ E_1^-=2k-2m+1, \nonumber \\
E_n^-=n+2k, \ n=2,3,4,...
\end{eqnarray} 
In this special case the potential
energy $V_-(x)$ coincides with the one studied in \cite{23,24}.

\section{CES potentials}

In this section we develop a subsequent method for constructing
the CES potentials using the results of the previous section.

Suppose that $W_1(x)$ is a given superpotential that corresponds
to the exactly solvable potential $V_-^{(1)}(x)$. The example
of such superpotential is a shape invariant one \cite{22}.
As a result of (\ref{171}) $V_+(x)$ is also exactly solvable and thus
for $H_+$ we know in explicit form all energy levels
and the corresponding eigenfunctions.
Then using SUSY transformations (\ref{11}), (\ref{12}) we can easily
calculate all excited energy levels and wave functions of its SUSY
partner $H_-$, the wave function of the ground state is given
by (\ref{9}). 
But to do this we must have the superpotential $W(x)$ which is 
expressed through $\phi(x)$.
Because $W_1(x)$ is a given function it is convenient to represent
the superpotential $W(x)$ using (\ref{33}) and (\ref{34}) in the following
form
\begin{equation}
W(x)=W_1(x)+{\phi''(x)\over \phi'(x)}.
\end{equation}
Similarly, the new exactly solvable potential $V_-(x)$ can be written
as follows
\begin{equation} \label{Vm}
V_-(x)={1\over 2}(W_1^2(x) + W'_1(x))+
\left({\phi''(x)\over \phi'(x)}\right)^2+
2W_1(x){\phi''(x)\over \phi'(x)}-\epsilon.
\end{equation}
In this expressions the function $\phi(x)$ is not arbitrary one
but must satisfy (\ref{34}) for a given $W_1(x)$.
Thus we must solve equation (\ref{34}) with respect to
$\phi(x)$ for a given $W_1(x)$ which can be written in the
following form
\begin{equation} \label{Ephi}
{1\over 2}\phi''(x) +W_1(x)\phi'(x)=\epsilon \phi(x).
\end{equation}

In order to transform this equation to Schr\"{o}dinger type
equation
let us write $\phi(x)$ in the form 
\begin{equation} \label{f}
\phi(x)= f(x) \exp\left(-\int dx W_1(x)\right).
\end{equation}
The new function $f(x)$ satisfies the equation
which can be rewritten
as follows
\begin{equation} \label{Ef}
-{1\over 2}f''(x) + V_+^{(1)}(x)f(x)=-\epsilon f(x),
\end{equation}
where
\begin{equation}
V_+^{(1)}(x)={1\over 2}(W_1^2(x)+W'_1(x)).
\end{equation}
As we see it is a Schr\"{o}dinger type equation of SUSY quantum
mechanics but with negative energy.
The sign in the right hand side of equation (\ref{Ef}) can be changed
using a duality transformation called also as anti-isospectral
transformation \cite{15}
\begin{equation}
\xi = ix.
\end{equation} 
Then equation (\ref{Ef}) reads
\begin{equation} \label{Etf}
-{1\over 2}{d^2 {\tilde f(\xi)}\over d\xi^2}+\tilde V_-^{(1)}(\xi)
{\tilde f(\xi)}=\epsilon {\tilde f(\xi)},
\end{equation}
where we have introduced the notations
\begin{eqnarray}
{\tilde f(\xi)}=f(-i\xi), \label{tf} \\
\tilde V_-^{(1)}(\xi)=-V_+^{(1)}(-i\xi)
={1\over 2}
\left( {\tilde W}_1^2(\xi )-{d {\tilde W}_1(\xi)\over d\xi}\right), \\
{\tilde W}_1(\xi)=i W_1(-i\xi).\label{tW}
\end{eqnarray}
In the present paper we shall consider only such superpotentials
$W_1(x)$ for which  
the dual superpotential ${\tilde W}_1(\xi)$ 
is real function of $\xi$.
Then equation (\ref{Etf}) is ordinary Shr\"{o}dinger equation
of SUSY quantum mechanics.
Using (\ref{f}) and (\ref{tf}) the solutions of equation (\ref{Ephi})
can be expressed via the solutions of equation (\ref{Etf})
in the following form
\begin{equation} \label{phi}
\phi(x)=\tilde f(\xi)\exp\left(\int d\xi \tilde W_1(\xi)\right)
={\tilde f(ix)\over \tilde f_0(ix)},
\end{equation}
where $\tilde f(\xi)$ is the solution of equation (\ref{Etf})
for the energy $\epsilon$, 
$f_0(\xi)$ is the solution of the same equation for the zero energy
$\epsilon=0$.
Note, that for obtaining $\phi(x)$ it is not necessary to use
only square integrable solutions of equation (\ref{Etf}). 
The solutions
must be such that $\phi(x)$ is a monotonic function that has one node. 

Thus, now we have the problem to solve equation (\ref{Etf}).
In order to solve this equation the following fact is important.
If $W_1(x)$ is such that the corresponding SUSY partners 
$V_{\pm}^{(1)}(x)$ belong to the class of shape invariant potentials
then
$\tilde W_1(\xi)$ gives also the shape invariant SUSY partners
$\tilde V_{\pm}^{(1)}(\xi)$.
To see this recall that the superpotential in the 
shape invariant case satisfies the following equation \cite{22}
\begin{equation} \label{SiW}
W^2_1(x,\alpha)+{dW_1(x,\alpha)\over dx}=
W^2_1(x,\alpha_1)-{dW_1(x,\alpha_1)\over dx}+2R,
\end{equation}
where 
the superpotential $W_1(x)$ in the left and right hand sides of this 
equation have different values of parameters
$\alpha$ and $\alpha_1$, 
the remainder $R$ does not depend on $x$. 
From this equation using the definition (\ref{tW})
we obtain
\begin{equation} \label{SitW}
\tilde W^2_1(\xi,\alpha_1)+{d\tilde W_1(\xi,\alpha_1)\over d\xi}=
\tilde W^2_1(\xi,\alpha)-{d\tilde W_1(\xi,\alpha)\over d\xi}+2R.
\end{equation}
As we see $\tilde W_1(\xi)$ also satisfies the shape invariant equation.
Note, that in compare to (\ref{SiW}) 
the set of parameters $\alpha$ in the equation (\ref{SitW})
is replaced by $\alpha_1$
and vice versa.

Thus $\tilde W_1(\xi)$ gives the shape invariant SUSY partners
$\tilde V_\pm^{(1)}(\xi)$ and equation
(\ref{Etf}) can be solved exactly. 
Using these solutions on the basis of (\ref{phi})
we get $\phi (x)$. 
Substituting $\phi(x)$ into (\ref{Vm})
we obtain a new exactly solvable potential $V_-(x)$
which is lower SUSY partner of the known
exactly solvable potential $V_+(x)$.
Of course we must verify that $\phi(x)$ satisfies the considered
above conditions, namely, $\phi(x)$ must be monotonic
function with one node.

In conclusion of this section let us consider some examples.

\subsection{Example 3}

Let us put
\begin{equation}
W_1(x)= x.
\end{equation}
Such a choice corresponds to linear harmonic oscillator.
The dual superpotential in this case has the same form
as $W_1(x)$
\begin{equation}
\tilde W_1(\xi)= \xi
\end{equation}
and thus equation (\ref{Etf}) 
is the Shr\"{o}dinger equation for linear harmonic
oscillator. Then using the well-known wave functions of stationary
states of linear harmonic oscillator on the basis of (\ref{phi}) 
we obtain
$\phi (x)=\beta H_{2k+1}(ix)$ ($\beta$ is some constant),
which is exactly equal to (\ref{E1phi}). Here each $k$
corresponds to the energy $\epsilon=2k+1$.
Note, that in order to satisfy appropriate conditions for $\phi(x)$
we select only odd solutions of equation (\ref{Etf}).

This example reproduces the result of the first example
of section 3 in  the special case $ \gamma =1$.

\subsection{Example 4}

Let us consider a superpotential $W_1(x)$ 
that corresponds to the Rosen-Morse oscillator.
In this case 
\begin{equation}
W_1(x)=\alpha \tanh (x)
\end{equation}
is shape invariant. The dual superpotential
\begin{equation} \label{E4tW1}
\tilde W_1(\xi)=\alpha \tan (\xi)
\end{equation}
is also shape invariant. The potential energy corresponding to
(\ref{E4tW1}) reads
\begin{equation} \label{E4V}
\tilde V_-^{(1)}(\xi)= {\alpha (\alpha -1) \over 2 \cos^2(\xi)}-
{\alpha^2 \over 2}.
\end{equation}
As we see the dual potential energy has singularities at
the points
\begin{equation} \label{E4xi}
\xi_n={\pi\over 2}+\pi n, \ \ n=0, \pm 1, \pm 2,....
\end{equation}
Traditionally, Schr\"{o}dinger 
equation (\ref{Etf}) with potential energy (\ref{E4V})
is considered on the interval between two neighbouring singularities 
using zero boundary conditions for solutions $\tilde f(\xi)$.
Without any problem we consider the solutions of (\ref{Etf})
on the full $\xi$-line which take zero value in all points (\ref{E4xi})
\begin{equation}
\tilde f(\xi_n)=0, \ \ n=0, \pm 1, \pm 2, ....
\end{equation}
Such solutions can be easily obtained with the help of SUSY
quantum mechanics. Using three first odd solutions of (\ref{Etf}) for
$\phi(x)$ given by (\ref{phi}) we obtain
\begin{eqnarray}
\phi_1(x)=\sinh (x), \ \ for \ 
\epsilon=\epsilon_1  
\label{E4phi1} \\
\phi_3(x)=[1-\alpha+(2+\alpha )\cosh(2x)]\sinh x, 
\label{E4phi3} 
for \  \epsilon =\epsilon_3, \\
\phi_5(x)=[6+\alpha+3\alpha^2-4(\alpha^2+2\alpha-3)\cosh(2x)+
\label{E4phi5} \nonumber\\
(\alpha^2+7\alpha+12)\cosh(4x)]\sinh x,  
for \ \ \epsilon =\epsilon_5,
\end{eqnarray}
where
\begin{equation} \label{E4epsk}
\epsilon_k =((\alpha +k)^2-\alpha^2)/2.
\end{equation}
Note that we may directly verify that 
functions $\phi(x)$ given by 
(\ref{E4phi1}), (\ref{E4phi3}), (\ref{E4phi5}) indeed satisfy 
equation (\ref{Ephi}). 
Substituting the obtained 
$\phi_k(x)$ and $\epsilon=\epsilon_k$ into (\ref{Vm}) we get the set of
exactly solvable potentials $V_-(x,k)$
(here we explicitly write down the dependence of the potential on $k$)  
\begin{equation} \label{VN}
V_-(x,k)=\tanh^2 (x)
\left({1\over2}\alpha (\alpha -1)+\Phi_k(x)(\Phi_k(x)+2\alpha)\right)
-\epsilon_k +{\alpha\over2},
\end{equation} 
where
\begin{eqnarray*}
\Phi_1(x)=1, \\
\Phi_3(x)={3(2+\alpha)\cosh(2x)+\alpha+3 \over 
(2+\alpha)\cosh(2x)-\alpha-1}, \\ 
\Phi_5(x)={(3+\alpha)[5(4+\alpha)\cosh(4x)+4(5-\alpha)\cosh(2x)]+
\alpha (5-\alpha )+30\over
(3+\alpha)[(4+\alpha)\cosh(4x)-4(1+\alpha)\cosh(2x)]+
3(1+\alpha)(2+\alpha)}. 
\end{eqnarray*}
The potential $V_-(x,1)$ reproduces the Rosen-Morse one.
Other potentials are new exactly 
solvable ones. 

Note that for $k=3$ the potential can be written
in the following explicit form
\begin{eqnarray}
V_-(x,3)=
-{4(3+2\alpha)\over ((2+\alpha)\cosh(2x)-1-\alpha)^2}+
{4(1+\alpha)\over(2+\alpha)\cosh(2x)-1-\alpha} \\ \nonumber
-{(1+\alpha)(2+\alpha)\over 2 \cosh^2 x}+
{(3+\alpha)^2\over 2}
\end{eqnarray}
and was previously obtained by us in \cite{25}. 

Note, that $\epsilon$ is the
parameter of the superpotential and thus the parameter of the potentials
$V_{\pm}(x)$. 
As we see $V_-(x)$ is exactly solvable when $\epsilon$ is equal
to a certain fixed value. 
It is worth to stress that for given functions $\phi(x)$ 
(for example (\ref{E4phi3}), (\ref{E4phi5}))
and arbitrary $\epsilon$ it is always possible 
using the results of previous section
to construct
QES potentials with two known eigenstates.
These QES potentials at certain fixed values of $\epsilon$
(\ref{E4epsk}) become exactly solvable (\ref{VN})
and can be treated as
CES potentials.

\section{Conclusions}

We have developed a new SUSY method for constructing QES potentials 
for which we know in explicit form energy levels and wave functions
of the ground and first excited states.
From the obtained general expressions for QES potential and
wave functions of the ground and first excited states
the following interesting fact can be derived.
The ratio of the wave functions of first excited state and 
ground state and the distance between corresponding 
energy levels entirely
determine the potential energy.

The method developed for constructing QES potentials
with two known eigenstates
is extended for generating CES potentials
which are exactly solvable at certain fixed values of parameter
$\epsilon$. 
Finally, this new exactly solvable potential is the lower SUSY partner
to the known exactly solvable potential.
In this sense our method for generating CES potentials is similar 
to the method proposed in \cite{19,20} although the realization 
is different. In addition, our method gives the interesting
relation between QES and CES potentials.
Namely, when the parameter $\epsilon$ of QES potentials is equal
to a certain fixed values then QES potentials
become exactly solvable and can be treated as CES ones.

Note, that important moment in our approach for 
generating CES potentials is the duality transformation which
we use to transform equation (\ref{Ephi}) 
to Schr\"{o}dinger type equation.
In the present paper we consider only superpotential for which 
the dual one (\ref{tW}) is a real function of new variable.
The case of complex dual superpotential is more
complicated and we plan to consider this case in future.
It will give a possibility to extend the class of
CES potentials which can be obtained by 
suggested in the present paper method.
 
\pagebreak


\begin{thebibliography}{25}
\bibitem{1} V.~Singh, S.~N.~Biswas, K.~Dutta, Phys. Rev. D {\bf 18} 
            (1978) 1901.
\bibitem{2} G.~P.~Flessas, Phys. Lett. A {\bf 72} (1979) 289.
\bibitem{3} M.~Razavy, Am. J. Phys. {\bf 48} (1980) 285; Phys. Lett A 
{\bf 82} (1981) 7.
\bibitem{4} A.~Khare, Phys. Lett. A {\bf 83} (1981) 237.
\bibitem{5} A.~V.~Turbiner, A.~G.~Ushveridze, Phys. Lett. A {\bf 126} 
             (1987) 181.
\bibitem{6} A.~V.~Turbiner, Commun. Math. Phys. {\bf 118} (1988) 467.
\bibitem{7} M.~A.~Shifman, Int. Jour. Mod. Phys. A {\bf 4} (1989) 2897.

\bibitem{8} O.~B.~Zaslavskii, V.~V.~Ul'yanov, V.~M.~Tsukernik,
             Fiz. Nizk. Temp. {\bf 9} (1983) 511.
\bibitem{9} O.~B.~Zaslavsky, V.~V.~Ulyanov, Zh. Eksp. Teor. Fiz.
             {\bf 87} (1984) 1724.
\bibitem{10} D.~P.~Jatkar, C.~Nagaraja Kumar, A.~Khare, Phys. Lett. A 
             {\bf 142} (1989) 200.
\bibitem{11} P.~Roy, Y.~P.~Varshni, Mod. Phys. Lett. A {\bf 6}
             (1991) 1257.
\bibitem{12} A.~Gangopadhyaya, A.~Khare, U.~P.~Sukhatme, 
             Phys. Lett. A {\bf 208} (1995) 261.
\bibitem{13} V.~V.~Ulyanov, O.~B.~Zaslavskii, J.~V.~Vasilevskaya,
            Fiz. Nizk. Temp. {\bf 23} (1997) 110. 
\bibitem{14} A.~G.~Ushveridze, Quasi-exactly solvable models in quantum 
            mechanics,
            Institute of Physics Publishing, Bristol (1994).
\bibitem{15} A.~Krajewska, A.~Ushveridze and Z.~Walczak,
             Mod. Phys. Lett. {\bf A 12} (1997) 1225.
\bibitem{16} A.~Khare, B.~P.~Mandal, 
             J. Math. Phys. {\bf 39} (1998) 3476.
\bibitem{21} A. de Souza Dutra, Phys. Rev. A {\bf 47} (1993) R2435.
\bibitem{Nag} N.~Nag, R.~Roychoudhury, Y.~P.~Varchni, Phys. Rev. A
              {\bf 49} (1994) 5098.
\bibitem{Dutt} R.~ Dutt, A.~Khare, Y.~P.~Varchni, J. Phys. A:
              Math. Gen. {\bf 28} (1995) L107. 
\bibitem{19}  G.~Junker, P.~Roy, Phys. Lett. A {\bf 232} (1997) 155.
\bibitem{20} G.~Junker, P.~Roy, preprint quant-ph/9803024.
\bibitem{17} F.~Cooper, A.~Khare, U.~Sukhatme, Phys. Rep. {\bf 251} 
             (1995) 267.
\bibitem{18} G.~Junker, Supersymmetric methods in quantum and statistical
             physics (Springer, Berlin, 1996). 
\bibitem{25} V.~M.~Tkachuk,  Phys. Lett. A, {\bf 245} (1998) 177. 
\bibitem{26} V.~M.~Tkachuk, T.~V.~Kuliy, preprint quant-ph/9807025.
\bibitem{22} L.~E.~Gendenshteyn, Pisma Zh. Eksp. Teor.Fiz. {\bf 38}
             (1983) 299.
\bibitem{Shab} A.~Shabat, Inverse Prob. {\bf 8} (1992) 303.
\bibitem{Spir} V.~P.~Spiridonov, Phus. Rev. Lett. {\bf 69} (1992) 398.
\bibitem{SS} D.~T.~Barclay, R.~Dutt, A.~Gangopadhyaya, A.~Khare,
             A.~Pagnamenta, U.~Sukhatme,
             Phys. Rev. A {\bf 48} (1993) 2786. 
\bibitem{23} V.~G.~Bagrov, B.~F.~Samsonov, Teor. Mat. Fiz. {\bf 104}
             (1995) 356.
\bibitem{24} V.~G.~Bagrov, B.~F.~Samsonov, J. Phys. A {\bf 29}
             (1996) 1011.               
\end{thebibliography}
\end{document}